\documentstyle[12pt,aasms4]{article}

\begin{document}

\title{The nuclear cusp slopes of dwarf elliptical galaxies
\footnote{Based on observations with the NASA/ESA {\it Hubble Space
Telescope} - GO 6352, obtained at the Space Telescope Science
Institute, which is operated by the Association of Universities for
Research in Astronomy, Inc., under NASA contract No.  NAS5-26555}}

\author {Massimo Stiavelli}
\affil{Space Telescope Science Institute, 3700 San Martin Dr.,
Baltimore, MD 21218}

\author{Bryan W. Miller}
\affil{Gemini Observatory, Casilla 603, La Serena, Chile}

\author{Henry C. Ferguson, Jennifer Mack, and Bradley C. Whitmore}
\affil{Space Telescope Science Institute, 3700 San Martin Dr.,
Baltimore, MD 21218}

\and 
\author{Jennifer M. Lotz} 
\affil{Dept. of Physics and Astronomy, Johns Hopkins University, Baltimore, 
MD, 21218}

\begin{abstract}
We derive the light profiles for a sample of 25 dwarf elliptical
galaxies observed by us with HST/WFPC2 in F555W and F814W. These
profiles are fitted with Nuker, $R^{1/4}$, exponential, and Sersic
laws, and are also used to derive the nuclear cusp slopes $\gamma$. We
discuss the correlation of nuclear cusp slope with galactic
luminosity, presence of a nucleus, and type of light profile. The
results are compared to those found in the literature for elliptical
galaxies and the bulges of spiral galaxies. We find that as a class
the nuclear regions of dwarf ellipticals are very similar to those of
the exponential bulges of spiral galaxies, and have nuclear cusp
slopes shallower than those of bulges that were well fitted by a de
Vaucouleurs R$^{1/4}$ profile with the same luminosity. For the 14
nucleated galaxies in our sample this conclusion is less certain than
for the 11 non-nucleated objects since it relies on an extrapolation
of galaxy light under the nucleus.  In terms of their light profiles
and nuclear properties, most spheroidal stellar systems can be broadly
divided into two subclasses: the exponential shallow cusp objects and
the $R^{1/4}$, steep cusp objects. Membership of a class does not
appear to correlate with the presence of a massive stellar disk.

\end{abstract}

\section{Introduction}

Dwarf elliptical galaxies are the dominant galaxy population in
clusters of galaxies, yet we are still unsure about their formation
mechanism and their evolution. In addition, it is not clear whether
the nucleated and non-nucleated dwarfs share a common formation and
evolution differing only in their typical luminosity, or are separated
by more profound differences. In order to better understand the nature
of dwarf ellipticals, we have started a systematic study of the
properties of nearby dwarf ellipticals (hereafter dE) in the Virgo and
Fornax clusters and in the Leo group. We observed 25 dE galaxies with
the Wide Field and Planetary Camera 2 (hereafter WFPC2) onboard the
Hubble Space Telescope (HST). For each galaxy we obtained a visible
F555W and a near-IR F814W image. We have initially focussed our
attention on the properties of the globular cluster population in
dwarf ellipticals, and our results are presented in Miller et
al. (1998).  In the present paper we will instead study the nuclear
properties of these objects. The extended light profiles will also be
derived.

HST observations have deeply affected our view of the properties of
the nuclear regions of all types of galaxies. Expanding on early work
by Crane et al. (1993), Jaffe et al. (1994) and Lauer et al.  (1995)
showed that the inner regions of all early--type galaxies regular
enough to be studied, present a residual slope in their radial light
profiles down to the smallest spatial scales accessible to HST. No
galaxy presents a ``core'' i.e. a flat light profile in the inner
regions. Moreover, Lauer et al.  (1995) found that the inner cusp
slope $\langle \gamma \rangle \equiv -\langle d Log I(r)/d Log r
\rangle$ is correlated with global galactic properties: bright, boxy
ellipticals have the shallowest cusps, i.e., the lowest values of
$\gamma$, while fainter, disky ellipticals have the steepest cusps and
the highest values of $\gamma$.

The sample was later extended to include ellipticals with
kinematically decoupled nuclei (KDC, Carollo et al.  1997a,b) and the
bulges of spiral galaxies (Carollo \& Stiavelli 1998 and references
therein). While KDC were found indistinguishable in their nuclear
properties from regular ellipticals, the bulges of spirals showed a
bimodal behavior: those bulges that were well fitted, outside their
nuclei, by a de Vaucouleurs R$^{1/4}$ profile followed and extended to
lower luminosities the trend observed for ellipticals, with the
fainter systems being characterized by the steepest nuclear slopes. By
contrast, the large fraction of bulges well fitted by an exponential
light profile outside their nuclei showed a nuclear cusp slope much
shallower than that of R$^{1/4}$ bulges with the same
luminosity. Since the radial intervals used for determining the nature
of the light profile and for measuring the cusp slope were disjoint,
Carollo \& Stiavelli (1998) argued that this difference was intrinsic
and not a mere artifact of exponentials being flatter than R$^{1/4}$
profiles.  To prove this they introduced a modified exponential
becoming cuspy at small radii and found that no galaxy was well fitted
by such a profile and that, in particular, the central star clusters
were more compact than what could be obtained with such a cuspy
exponential.

If two classes of bulges exist, then a distinct possibility is that
dwarf ellipticals are also separated into two classes by their nuclear
properties and, perhaps, by their light profiles. The photometric
properties of dwarf galaxies have been studied in the past using
ground--based data (see, e.g., Binggeli and Cameron 1993, Vader and
Chaboyer 1994). These studies revealed the presence of a large variety
of light profile shapes in dwarf galaxies but were unable to reveal
their nuclear properties since their angular resolution was limited by
atmospheric seeing. This paper is devoted to exploring this issue by
exploiting high angular resolution HST photometry. In Section 2 we
briefly outline the data reduction and describe how the light profiles
were derived. A full description of the data reduction is given by
Miller et al. (2000). Section 3 addresses the light profile fits to
derive cusp slopes and profile type. The correlation of the derived
nuclear slopes with other galactic properties are presented and
discussed in Section 4.

\section{Observations and Data Reduction and Analysis}

For a detailed description of the observations and basic reduction we
refer to Miller et al.\ (1998, 2000). Our WFPC2 snapshot images are
taken in the F555W ($2 \times 230$ sec) and F814W (300 sec) bandpasses
with the galaxies centered in chip WF3.  The two F555W images are
combined and cleaned of cosmic rays (hereafter CR) using a program
kindly provided to us by Dr. Richard White. The program identifies
cosmic ray hits by sigma-clipping. The peak of each cosmic ray is
determined using a high threshold (usually 5 sigma) and the extended
area around the peak by using a lower threshold (usually 1 sigma). The
noise model includes the detector noise, Poisson noise and effects of
flat fielding.  The clean image is used as a template for identifying
cosmic rays in the single F814W image and removing them by
interpolation.

Isophotal profiles were derived on the images clean of CR by using the
{\tt ellipse} task in the STSDAS/isophote package. This task is based
on the software developed by R. I. Jedrzejewski (1987). Isophotes are
determined by least square fitting of the ellipse parameters.
Departures from purely elliptical isophotes are then computed by a
separate minimization once the best ellipses have been determined. The
fit was carried out iteratively to mask any area affected by dust. In
particular, we first fitted the galaxy without any dust masking and
subtracted the resulting model to identify regions where dust was
present. Those regions were masked and the fit was repeated
again. Since dwarf galaxies are affected very little by dust, and
since we had previously tested extensively this procedure by
comparison with an independent fitting scheme (Carollo et al. 1997c),
we felt that it was not necessary for these objects to carry out
isophotal fits on every image using two different algorithms. The
isophotal fits were carried out independently on the combined F555W
and the F814W images. The profiles in counts were calibrated to
magnitudes using the prescriptions by Holtzman et al. (1995), namely,
by iterative application of the equations:

\begin{equation}
V = -2.5 \log{(DN_{F555W} s^{-1})}+21.725 -0.052 \times(V-I)+0.027 \times 
(V-I)^2+2.5 \times \log{2.006},
\end{equation}

\begin{equation}
I = -2.5 \log{(DN_{F814W} s^{-1})}+20.839 -0.062 \times (V-I)+0.025 \times 
(V-I)^2+2.5 \times \log{2.006}
\end{equation}

The iterations are started by assuming as initial value of $V-I$ the
one derived by the above equations for $V-I = 0$. Three iterations
were typically necessary to achieve convergence. Values of surface
brightness are derived from the magnitudes by correcting for the pixel
area ($0\farcs1 \times 0\farcs1$) and for foreground extinction with
$A_V = 0.75 A_B$ and $A_I = 0.44 A_B$.  The light and color profiles
for the sample galaxies are shown, respectively, in Figures 1 and 2.
Note that the observed color profiles in Figure 2 are obtained from
direct subtraction of the profiles in magnitudes arcsec$^{-2}$ and as
such are not corrected for the different PSF in F555W and F814W nor
for the presence of a nucleus or a globular cluster close to the
center.

The apparent magnitudes for each galaxy were computed using two
different approaches. Firstly, we measured the flux from a 60$''$
aperture in the WF3 after sky and bright sources were removed and
replaced by a suitable galaxy flux obtained by interpolation.
Secondly, we integrated the flux in the best fit to the galaxy surface
brightness profile (see Section 3) by extrapolating beyond the last
measured point. The two methods have different advantages and
disadvantages. The aperture measurement is more sensitive to sky
subtraction errors. The method based on the integration of the surface
brightness profile is less sensitive to sky subtraction error but
suffers from the uncertainties of extrapolating the observed points.
We have adopted as apparent magnitude (listed in Table 1) the average
of these two measurements and as error the largest between the formal
error in the measurements (typically 0.1 mag) and the semi-difference
between the two measurements.

In addition to the apparent magnitude, Table 1 lists the adopted
extiction, the assumed distance for each of the target galaxies, and
their absolute magnitude.  The distances given in the table are
identical to those used in Miller et al.\ (1998), and correspond to
distance moduli of $(m-M)_0$ = 31.2 for the Virgo cluster, $(m-M)_0$ =
30.3 for the Leo group, and $(m-M)_0$ = 31.4 for the Fornax cluster.

In the following analysis we have adopted the definition of nucleated
dE of Miller et al. (1998, see also Table 2) who established the
presence of a nucleus simply by inspecting the images. This is
possible since dwarf galaxies tend to have especially flat core
profiles and rather compact nuclei which, when present, contrast very
well with the flat core.  Our isophotal fits confirm these
identification, i.e., no fainter nucleus has been identified in
galaxies considered to be non-nucleated by Miller et al. (1998).  Some
nuclei are off-center. For this reason, Miller et al. (1998)
determined for each galaxy the centering error $\sigma_{proj}$ and
required that a nucleus needed to be within $3 \sigma_{proj}$ from the
center. Moreover, a nucleus had to have a probability of more than 99
\% of not being a globular cluster seen in projection on the
nucleus. The off-center nuclear sources in VCC 1577 and VCC 1762 fail
the probability and the distance test, respectively, and are therefore
considered as globular clusters.

\section{Profile fits and cusp slopes}

The isophotal profiles have been fitted with a number of analytical
profiles following Carollo et al. (1997a,b).  The program carries out
a fit over a user specified radial interval while taking into account
the PSF characteristics of the data. This is done by convolving the
model with Tiny Tim (Krist 1999) PSFs obtained for F555W and F814W and
a G0 stellar spectrum. We have used both circularized PSFs obtained by
deriving the PSF profile and the original PSFs and found the same cusp
light profiles and slopes within the errors. Therefore most of the
fits have been carried out using the - faster - circularized PSF. Both
the model and the PSF are supersampled by a factor 5 (i.e. each pixel
is 0.020 arcsec) in order to avoid artifacts due to steep slopes near
the center. The model convolved with the PSF is then resampled and
compared to the data. The best-fit profile is thus unaffected by the
WFPC2 PSF. We have verified that our results do not depend on the
addition of a small (0.005 arcsec) jitter or on changing the stellar
spectrum used to generate the PSF to an F or K type; the change in
fitted parameters remaining smaller than the error bars.

A general problem faced when fitting light profiles of galaxies with a
nucleus is the uncertainty in separating galaxy and nuclear light.
For the dwarf galaxies in our sample the nucleus is fainter than the
galaxy by 3.7 - 7.4 magnitudes, with 5 magnitudes being a typical
value. Its overall contribution to the light is thus negligible even
though it dominates at very small radii. We have adopted the approach
of restricting our fits to those radii which appear unaffected by the
nucleus, i.e., typically outside 0.5 arcsec.  Although not nucleated
according to the Miller et al. (1998) definition, VCC 1577 and VCC
1762 have a bright source, likely a star cluster, close to their
center. We have been able to subtract this source. This was done
iteratively by modeling the source as small Gaussian convolved with
the PSF and by changing the total source flux and Gaussian width until
the nucleus could be subtracted without leaving either light excesses
or dips from the otherwise flat core. The results obtained by
repeating the fit for these ``clean'' galaxies agree within the errors
with those obtained by radially restricting the fit.

We first fitted the profiles with a ``Nuker law'' (Lauer et al.\ 1995,
Byun et al. 1996), namely:
\begin{equation}
I(R) = 2^{(\beta-\gamma)/\alpha} I_b \left(\frac{R_b}{R}\right)^\gamma
\left[1+\left(\frac{R}{R_b}\right)^\alpha\right]^{(\gamma-\beta)/\alpha},
\end{equation}
where $R$ is the projected radius, and $R_b$, $\alpha$, $\beta$,
$\gamma$, $I_b$ are fitted parameters. The cusp slope is derived in
two independent ways: either as the $\gamma$ parameter of the best
Nuker fit or as the mean slope between 0\farcs1 and 0\farcs5 of the
Nuker fit computed as $ \langle \gamma \rangle = -
\frac{\sqrt{\langle( \log{I} -\langle \log{I} \rangle)^2 \rangle}}{
\sqrt{\langle (\log{R} -\langle \log{R} \rangle)^2 \rangle}}$.  Since
our definition of $\langle \gamma \rangle$ relies on the use of the
best Nuker fit as an interpolating formula, it is corrected for the
WFPC2 PSF and for the possible presence of a central star cluster.
For objects without a nuclear star cluster, we checked that the
$\langle \gamma \rangle$ slope derived by fitting the whole profile or
the radially restricted profile were the same.  Generally, as observed
also by Byun et al.\ (1996) and Carollo et al.\ (1997), there is good
agreement between fitted value of $\gamma$ and the average of the
nuclear slope $\langle \gamma \rangle$. For a few objects, namely FCC
48, FCC 150, LGC 47, VCC 2029, the difference between the two
estimates is many times larger than the formal errors. However, in
these cases the values are close to zero.  In Table 2 we list the
Nuker fit parameters, including the slope $\gamma$, and the derived
value of the nuclear cusp slope $\langle \gamma \rangle$. The errors
listed are one sigma errors obtained by determining the change in each
parameter that increase the $\chi^2$ value of the fit by one. In the
following analysis we prefer to adopt the mean slope $\langle \gamma
\rangle$ but our results would not be changed by adopting the fitted
Nuker parameter $\gamma$.

To verify that the shallow values for the cusp slope that we obtain
are not purely an artifact of our fitting procedure, we repeated the
fit by considering for all galaxies (including those with a central
star cluster) the whole radial profile and by using a modified cuspy
exponential profile as the fitting function (see also Carollo \&
Stiavelli 1998).  This tests whether the central star clusters are
actually cusps within otherwise shallow cores. We find that in no case
does the cuspy exponential provide a good fit for meaningful values of
the fitted parameters.

Finally, we repeated our fits by using exponential profiles, R$^{1/4}$
profiles, and Sersic (1968) R$^{n}$ profiles. These fits were
typically performed outside 1'' radius.  The goodness of fit was
computed by estimating the value of $\chi^2$ per degree of
freedom. The error on the Sersic exponent has been found by requiring
the $\chi^2$ value to increase by one. Not surprisingly (see, e.g.,
Binggeli \& Cameron 1991 and Durrell 1997), we find that no galaxy in
our sample is well described by an R$^{1/4}$ law. The closest object
to an R$^{1/4}$ profile is VCC 1254 with a Sersic slope $n=0.29 \pm
0.01$. Out of 25 galaxies, 19 are well fitted by a simple exponential
profile, while 6 are described by Sersic profiles with $n$
significantly different from both 0.25 and 1 which are the values
giving an R$^{1/4}$ or an exponential, respectively (see Table
2). Generally, for those objects that are fitted equally well by an
exponential and a Sersic model, the two fits look very similar. The
one exception being FCC208 where a bump in the light profile is fitted
by the Sersic law and not by the exponential. Since the nucleus in
FCC208 departs from a point source, it is hard to establish whether
the feature should be considered part of the nucleus or part of the
galaxy. We think that it is likely that the bump arises from a PSF
feature and have chosen to consider it as part of the nucleus and
adopt the exponential fit. Clearly, adopting the Sersic fit for this
object would not have significantly changed our conclusions. If we
took the error on the Sersic exponent $n$ at face value, then for 13
out of 25 objects the exponent $n$ is different from one at the three
sigma level. However, in some of these cases, the exponent is close to
one so that the model profile remains essentially indistinguishable
from an exponential.  Note that for bulges of spiral galaxies Carollo
et al. (1997c) and Carollo \& Stiavelli (1998) found that most
galaxies regular enough to derive isophotal profiles could be
described by either exponentials or R$^{1/4}$ laws. Only a few bulges
could not be fitted by either and would presumably be fitted by a
Sersic law. The fitted dE profiles are shown in Figure 1 (solid line
for the best exponential fit and dashed line for the best Sersic fit)
and are used to obtain color profiles unaffected by central sources or
PSF dependencies shown in Figure 2.  The average color in the
0\farcs1-0\farcs5 region is given in Table 1.  In Figure 3 we plot the
histogram of the derived Sersic slopes for the sample identifying by
shading those objects that are also well fitted by an exponential
profile. As expected, the Sersic exponents cluster around unity for
those galaxies with light profiles also well fitted by exponential
profiles.  On the contrary, it is for those objects with Sersic slope
$n \simeq 0.5$ that neither an R$^{1/4}$ or an exponential profile
provide adequate fits and thus the extra parameter in the Sersic law
is essential to obtain a good fit.

\section{Results and Discussion}

As other galaxies observed with HST (Crane et al. 1993, Carollo et
al. 1997), dwarf galaxies are characterized by very shallow nuclear
color gradients. Some residual gradients can be seen in objects
containing a compact source near the nucleus but this is probably due
to an imperfect subtraction of the nuclear light.

Dwarf elliptical galaxies are generally characterized by shallow
nuclear cusp slopes. This conclusion is direct for non-nucleated
galaxies but relies on an extrapolation of the light profile below the
nucleus for nucleated objects. While the uncertainties in this
extrapolation for nucleated galaxies make this conclusion less robust
for these objects, the general similarity in derived properties for
nucleated and non-nucleated galaxies increases our confidence on this
result.

In Figure 4 we show the correlation of the average nuclear cusp slope
$\langle \gamma \rangle$ with the absolute magnitude $M_V$ for dEs and
other spheroidal stellar systems. In particular we include elliptical
and lenticular galaxies (Burstein et al. 1987; Bender et al. 1993) and
bulges of spiral galaxies (Carollo \& Stiavelli 1998). The dE objects
as a class appear to occupy roughly the same region of the $\langle
\gamma \rangle$-$M_V$ plane occupied by exponential bulges of spiral
galaxies. The nuclear slopes of dE galaxies are much shallower than
those of the low-luminosity end of the bright elliptical sequence.
Moreover, the few dEs characterized by steeper cusps are better fitted by
Sersic profiles (see also Table 2) rather than exponentials. There is
also an indication that dwarfs with a light profile well described by
an exponential profile have somewhat shallower nuclear slopes than
exponential bulges of the same luminosity. 

The location of a galaxy in the $<\gamma>$-$M_V$ plane appears to be
mildly influenced by the nucleated/non--nucleated nature of the dE
(Figure 5).  Indeed, by adopting the definition of nucleated dE of
Miller et al. (1998, see also Table 2), we find that nucleated
galaxies have systematically steeper slopes than non-nucleated
ones. However, a decrease in the cusp slopes of nucleated galaxies by
only 0.1 would produce a significant overlap between nucleated and
non-nucleated objects.  Unfortunately, we cannot exclude with
confidence the possibility that the higher cusp slopes measured in
nucleated dwarfs might be due to a systematic effect in separating
galaxy light from that of the nucleus. If part of the effect is real
rather than an artifact of the fitting, this would agree with the
indication that exponential bulges of spirals, that are all nucleated,
have a somewhat steeper light profiles than non-nucleated dwarf
ellipticals.  However, one of the steepest cusps is observed in VCC
917 which is non nucleated (but has a Sersic profile).

We have verified that neither the nuclear slope $\langle \gamma
\rangle$ nor the profile type (exponential or Sersic) seem to
correlate with the specific globular cluster frequency $S_N$ (as given
by Miller et al. 1998). Similarly, we do not see any correlation between
cusp slope and cusp color.

Our findings demonstrate that the anti-correlation between nuclear
slope and spheroid luminosity seen in bright elliptical galaxies
breaks down for dwarf ellipticals. With regard to their nuclear
properties dwarf ellipticals are more closely related to the
exponential bulges of spiral galaxies than to bright elliptical
galaxies. This is particularly interesting in light of the possibility
that some dwarfs ellipticals may actually be the bulges of stripped
spirals (Kormendy 1985). Regardless of the specific formation
processes, it appears that spheroidal stellar systems can be broadly
subdivided into two classes sharing the same light profile and nuclear
properties regardless of the presence of a surrounding stellar disk.

\acknowledgements

We thank Marcella Carollo and Tim de Zeeuw for discussions and an
anonymous referee for comments that helped improve the paper. MS
thanks the Lorentz Center of Leiden University where some of this work
was carried out, the European Space Agency for support and the Scuola
Normale Superiore of Pisa for hospitality. Support for this work was
provided by the STScI DDRF grant D001.82173 and NASA through grant
number GO-06532.01-95A from the Space Telescope Science Institute,
which is operated by AURA under NASA contract NAS5-26555.

\clearpage
%\bibliography{ajmnemonic,/home/bmiller/tex/refs/bib}
%\bibliographystyle{aj}

\clearpage
\figcaption[stiavelli.fig1a.ps]{Light profiles for the sample
galaxies. Only about 20 points are shown to avoid clutter. The solid
lines are the best exponential fits while the dashed lines are the
best Sersic fits to the light profiles. For some galaxies the two
curves are essentially identical. The adopted model is indicated in
each panel below the galaxy name.}

\figcaption[stiavelli.fig2a.ps]{Color profiles for the sample galaxies. The
observed V-I profiles inside 1'' are affected by compact sources
at or near the center which may have different colors from the
underlying population. The solid lines are the colors derived from the
independent exponential or Sersic fit to the V and I profiles and
therefore are corrected for PSF and do not include the effect of any
central source. Nuclear color gradients appear to be negligible for
all the objects.}

\figcaption[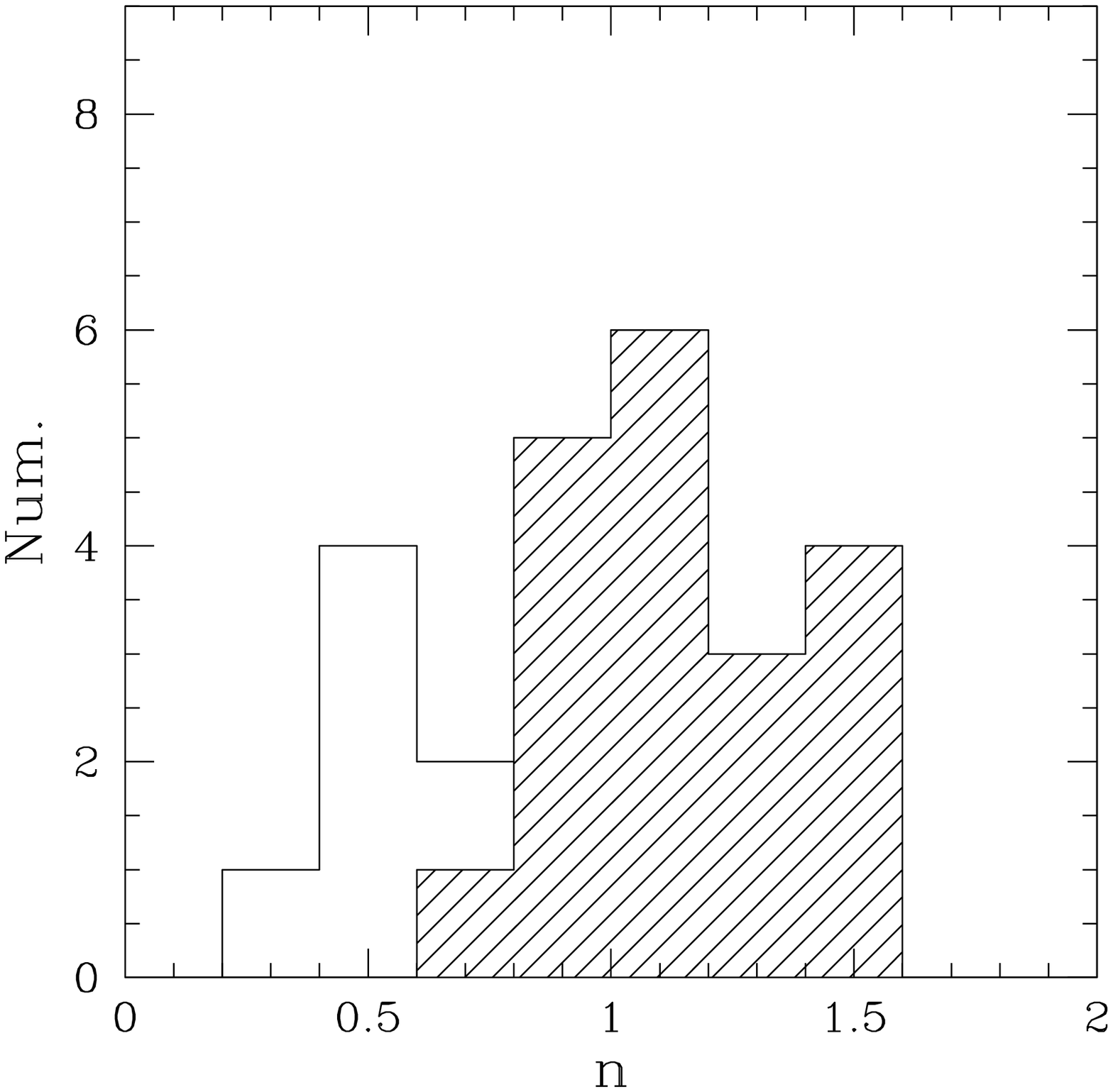]{Histogram of the values of the Sersic
parameter $n$ derived from the light profile fits. The shaded
histogram gives the values for those galaxies with an acceptable
exponential fit. No galaxy in our sample is well fitted by an
$R^{1/4}$ profile and 6 out of 25 are well fitted by a Sersic profile
with slope intermediate between $n$=0.25 and $n$=1.}

\figcaption[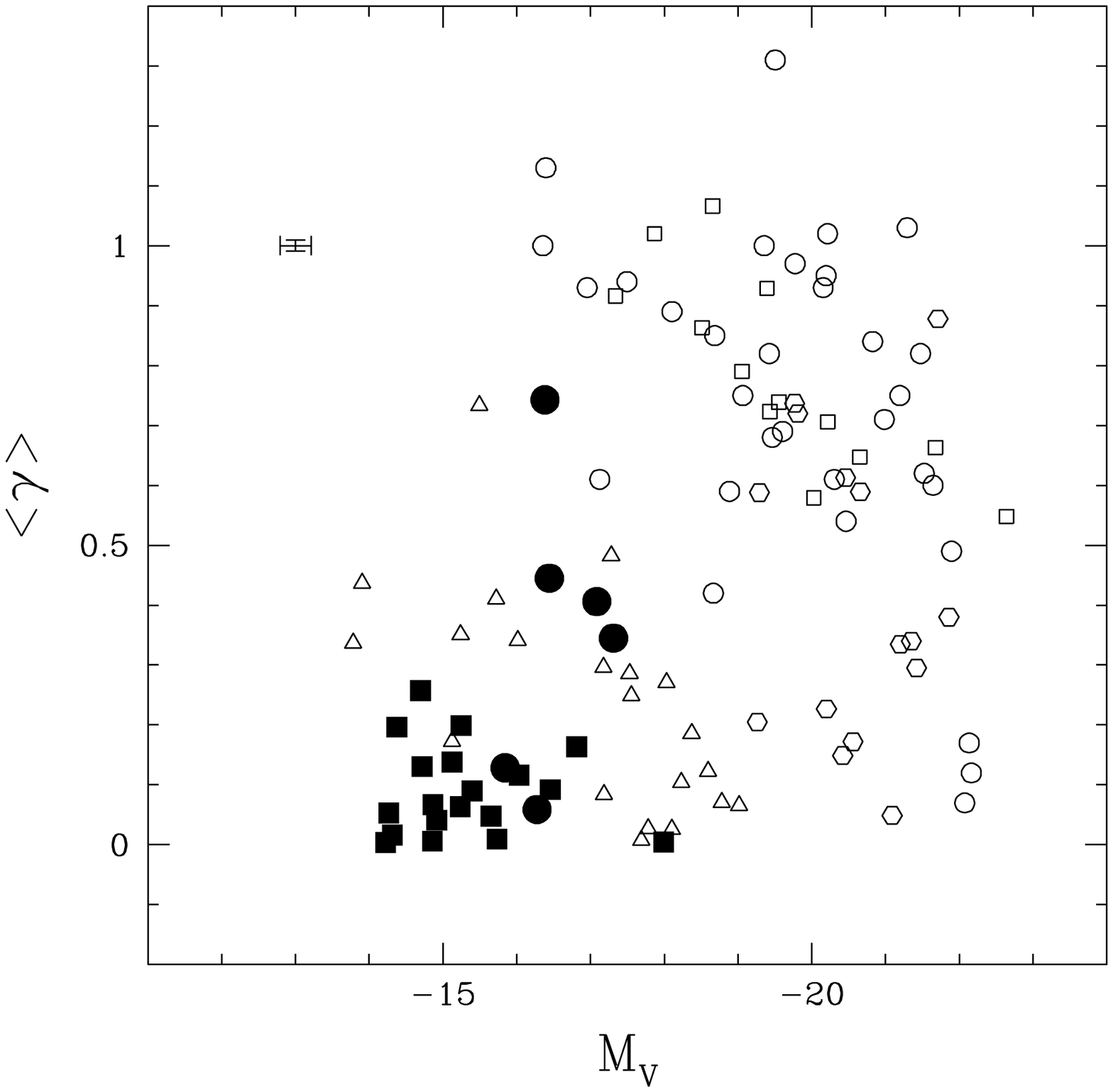]{The correlation between nuclear slope
$\langle \gamma \rangle$ and absolute magnitude $M_V$ is shown for the
dwarf galaxies in our sample (filled symbols) and other objects from
the literature. Dwarf ellipticals with exponential light profiles are
plotted as filled squares, those with Sersic profiles as filled
circles. Open triangles are the exponential bulges of Carollo et
al. (1998), the open circles are the $R^{1/4}$ bulges from Carollo et
al. (1998). The squares are elliptical galaxies from Lauer et
al. (1995) and the hexagons ellipticals from Carollo et
al. (1997). The error bars in the upper left corner represent the
median error.}

\figcaption[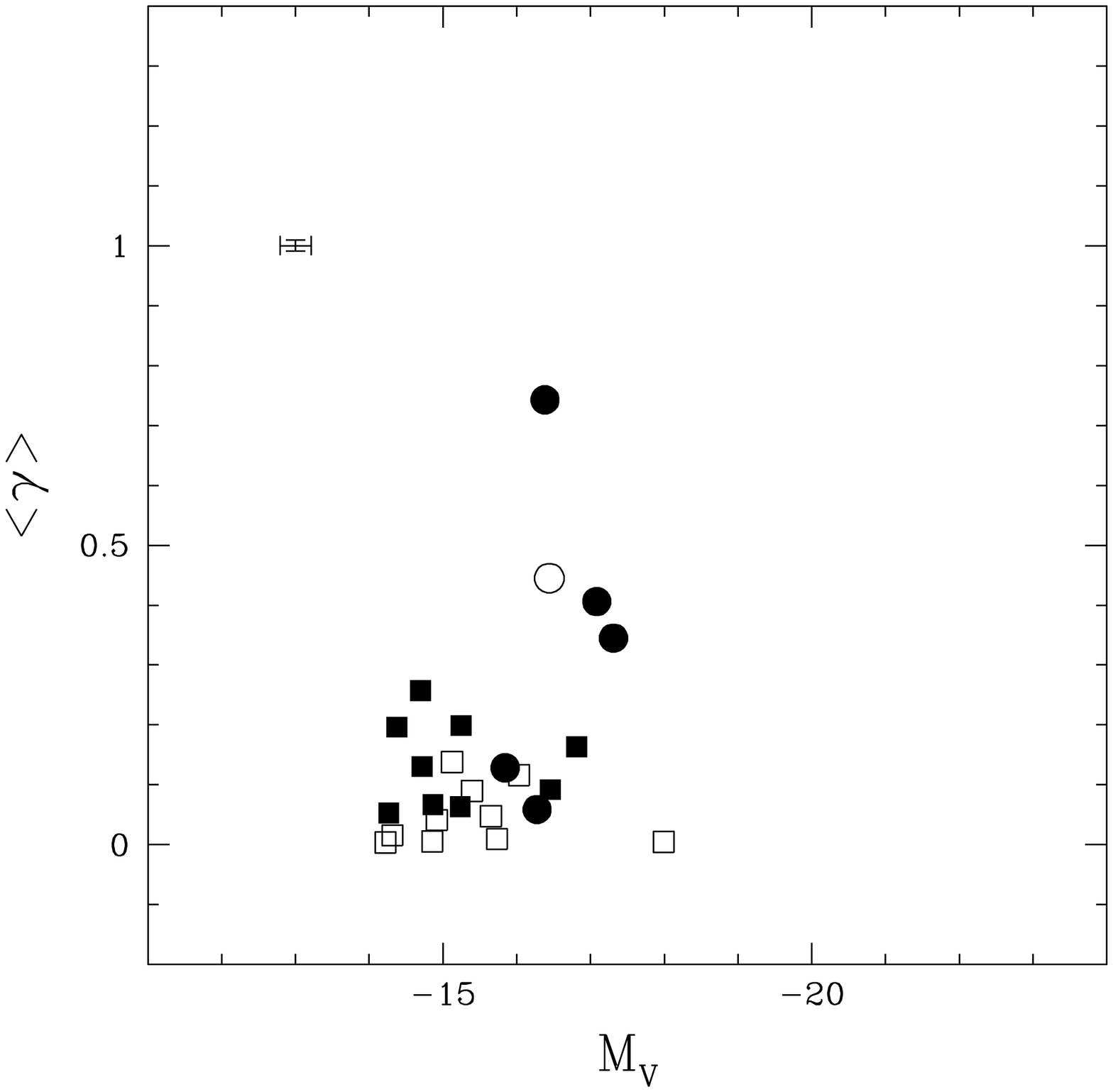]{We show the effect of the presence of a
nucleus on the nuclear slope by plotting nucleated dwarf ellipticals
(filled symbols) and non-nucleated dwarfs (open symbols) on the
$<\gamma>$-$M_V$ plane. Galaxies with a Sersic profile are
represented by circles, those with an exponential profile by squares.
The error bars in the upper left corner represent the
median error.}

\end{document}